\documentclass[aps,prl,preprint,superscriptaddress,amssymb,floatfix]{revtex4-1}
\usepackage{color}
\usepackage{afterpage}
\usepackage{pdfpages}
\usepackage{graphicx}

\begin{document}
\title{Reorientation of the bicollinear antiferromagnetic structure at the surface of Fe$_{1+y}$Te bulk and thin films}
\author{Torben H\"anke}
\email[Corresponding author: ]{thaenke@physnet.uni-hamburg.de}
\affiliation{Department of Physics, Hamburg University, 20355 Hamburg, Germany}
\author{Udai Raj Singh}
\affiliation{Department of Physics, Hamburg University, 20355 Hamburg, Germany}
\author{Lasse Cornils}
\affiliation{Department of Physics, Hamburg University, 20355 Hamburg, Germany}
\author{Sujit Manna}
\affiliation{Department of Physics, Hamburg University, 20355 Hamburg, Germany}
\author{Anand Kamlapure}
\affiliation{Department of Physics, Hamburg University, 20355 Hamburg, Germany}
\author{Martin Bremholm}
\affiliation{Center for Materials Crystallography, Department of Chemistry and iNANO, Aarhus
University, DK-8000 Aarhus C, Denmark}
\author{Ellen Marie Jensen Hedegaard}
\affiliation{Center for Materials Crystallography, Department of Chemistry and iNANO, Aarhus
University, DK-8000 Aarhus C, Denmark}
\author{Bo Brummerstedt Iversen}
\affiliation{Center for Materials Crystallography, Department of Chemistry and iNANO, Aarhus
University, DK-8000 Aarhus C, Denmark}
\author{Philip Hofmann}
\affiliation{Department of Physics and Astronomy, Interdisciplinary Nanoscience Center, Aarhus
University, DK-8000 Aarhus C, Denmark}
\author{Jin Hu}
\affiliation{Department of Physics and Engineering Physics, Tulane University, New Orleans, LA 70118, USA}
\author{Jens Wiebe}
\email[Corresponding author: ]{jwiebe@physnet.uni-hamburg.de}
\affiliation{Department of Physics, Hamburg University, 20355 Hamburg, Germany}
\author{Zhiqiang Mao}
\affiliation{Department of Physics and Engineering Physics, Tulane University, New Orleans, LA 70118, USA}
\author{Roland Wiesendanger}
\affiliation{Department of Physics, Hamburg University, 20355 Hamburg, Germany}
\begin{abstract}
\textbf{Establishing the relation between the ubiquitous antiferromagnetism in the non-superconducting parent compounds of unconventional superconductors and their superconducting phase is believed to be important for the understanding of the complex physics in these materials. Going from the bulk systems to thin films strongly affects the phase diagram of unconventional superconductors. For Fe$_{1+y}$Te, the parent compound of the Fe$_{1+y}$Se$_{1-x}$Te$_x$ superconductors, bulk sensitive neutron diffraction has revealed an in-plane oriented bicollinear antiferromagnetic structure. Here, we show by spin-resolved scanning tunneling microscopy that on the surfaces of bulk Fe$_{1+y}$Te, as well as on thin films grown on the topological insulator Bi$_2$Te$_3$, the spin direction is canted both away from the surface plane and from the high-symmetry directions of the surface unit cell, while keeping the bicollinear magnetic structure. Our results demonstrate that the magnetism at the Fe-chalcogenide surface markedly deviates from a simple in-plane oriented bicollinear antiferromagnetic structure, which implies that the pairing at the surface of the related superconducting compounds might be different from that in the bulk.}
\end{abstract}
\maketitle
The physics of many transition metal oxides (TMO) is dominated by strong electronic correlations which leads to exotic ground states and excitations with a dominant role of electronic charge and spin degrees of freedom. For example, for cuprate based high temperature superconductors (HTSCs) there have been growing evidence that charge and spin density wave-like states are inherent to these materials with a significant impact on the corresponding excitation spectrum~\cite{tsa1995,t2004,hlk2004}. Indeed, spin and charge ordering appear to be a key feature for understanding the physics of HTSCs~\cite{Neto393,Comin390}. Furthermore, among the correlated electron systems the recently discovered iron-based superconductors~\cite{feas_xx} are of particular interest for the understanding of the interplay between superconductivity and magnetism. 
Especially, the iron-chalcogenide system Fe$_{1+y}$Se$_{1-x}$Te$_x$ has gained high interest due to the surprisingly high $T_{\text{C}}$ superconductivity in single layer films of FeSe~\cite{0256-307X-29-3-037402,FeUCARPES,nmat4153} and its unique interplay between magnetism and superconductivity\cite{nmat2800}. For bulk single crystals FeSe exhibits a superconducting transition temperature $T_{\text{C}}$ of no higher than 10~K~\cite{Hsu23092008} but in a single unit cell (UC) thick film grown on SrTiO$_3$ $T_{\text{C}}$ can be increased above 100~K~\cite{nmat4153}. This finding has spurred numerous investigations aiming at the understanding of how superconductivity evolves in transition metal chalcogenides from bulk to ultra-thin films. In particular, scanning tunneling microscopy (STM) with atomic-scale resolution has proven to be an indispensable tool for revealing the real-space electronic structure~\cite{Kohsaka1380_,Pasupathy196}. However, most of the recent STM studies on TMO and iron-based superconductors have been focusing only on the charge degrees of freedom. With the recent developments in spin-polarized STM (SP-STM)~\cite{RevModPhys.81.1495,oswald}, which accesses both charge and spin degrees of freedom on the atomic length scale, detailed investigations of TMO compounds and iron-based superconductors have now become possible.

Since Fe$_{1+y}$Te exhibits double-stripe (bicollinear) antiferromagnetic (AFM) order, contrasted with single-stripe AFM order in iron pnictide superconductor parent compounds, the study on its mechanism of magnetic order has recently attracted a lot of attention. It provides a nonpolar charge-neutral Te terminated surface upon cleaving~\cite{PhysRevB.83.220502,PhysRevB.87.214508,Sugimoto201385}. Furthermore Fe$_{1+y}$Te can be grown \textit{in situ} by molecular beam epitaxy (MBE) with high quality~\cite{PhysRevB.91.220503,PhysRevB.93.041101}. In bulk, Fe$_{1+y}$Te is the nonsuperconducting parent compound of the transition metal chalcogenide system Fe$_{1+y}$Se$_{1-x}$Te$_x$\cite{nmat2800} and exhibits an AFM bicollinear magnetic structure below its N\'{e}el-temperature $T_{\text{N}}$~\cite{Fuchart,PhysRevLett.102.247001,PhysRevB.79.054503}. Depending on its excess iron concentration $y$ the N\'{e}el-temperature varies from $T_{\text{N}}\approx60\text{-}70$~K~\cite{PhysRevB.88.094509}. Using neutron diffraction, it has been shown that the bulk Fe spins are pointing along the diagonal of the Fe-Fe square network~\cite{PhysRevLett.102.247001,PhysRevB.79.054503}. The crystal structure and corresponding bicollinear AFM order are schematically shown in Figs.~\ref{fig:FeTe_simple_topo}(a,b). The magnetic phase transition of bulk Fe$_{1+y}$Te is accompanied by a structural phase transition, with the structure changing from a tetragonal to a monoclinic phase for which the lattice constant $a_{\text{Te}}$ is slightly larger than the lattice constant $b_{\text{Te}}$~\cite{JPSJ.79.102001,PhysRevLett.102.247001}. The bicollinear structure itself shows a commensurate AFM modulation along the $a_{\text{Te}}$-direction with a wave length of $\lambda_{\text{AFM}}=2a_{\text{Te}}$ and has a ferromagnetic (FM) coupling along the $b_{\text{Te}}$-direction~\cite{PhysRevLett.102.247001,PhysRevB.79.054503,PhysRevB.84.064403,PhysRevLett.102.247001}.

STM data obtained for Fe$_{1+y}$Te bulk and thin film samples have revealed atomic resolution of the Te-terminated surface and a superstructure on top of the atomic corrugation having a periodicity of $\lambda=2a_{\text{Te}}$~\cite{Sugimoto201385,PhysRevB.90.224503,Enayat08082014,udaj,PhysRevB.91.220503,PhysRevB.93.041101}. The observation of this additional $\lambda=2a_{\text{Te}}$ periodicity is mainly discussed in two contradictory models. On one hand, the interpretation for the $2a_{\text{Te}}$ superstructure is given in terms of a charge density wave (CDW), where the charge density of the top Te layer is modified by the spin density wave (SDW) of the underlying AFM order of the Fe-layer. The strong dependence on the bias voltage $V_{\text{bias}}$ suggests a complex interplay between the charge and the magnetic order~\cite{Sugimoto201385,PhysRevB.90.224503,PhysRevB.93.041101,machida1}. In contrast to a common representation of spin and charge modulations in correlated electron systems~\cite{PhysRevB.82.144522}, in this case the CDW would have the same periodicity as the SDW. Opposed to this interpretation, experiments performed with magnetically sensitive tips, which were prepared by attaching excess Fe atoms to the tip apex, give additional insight based on spin-polarized tunneling~\cite{Enayat08082014,udaj}. The $2a_{\text{Te}}$ superstructure can unambiguously be assigned to a direct imaging of the bicollinear antiferromagnetic order of the underlying Fe lattice. However, in previous studies the absolute orientation of the spins within the antiferromagnetic structure could not be revealed.

In this work we investigated the Fe$_{1+y}$Te surface of bulk samples and of thin Fe$_{1+y}$Te films grown on Bi$_2$Te$_3$ by spin-polarized STM, revealing the bicollinear antiferromagnetic structure in both sample systems. Moreover, by using Fe coated W-tips in a vector-magnet system, we were able to rotate the tip magnetization direction both within the surface plane as well as perpendicular to the surface. Since we thereby have access to the different components of the spin direction of the sample, we could unambiguously prove that the spin direction within the surface bicollinear antiferromagnetic structure is canted with respect to the crystallographic $b_{\text{Te}}$-direction.
\section{Results}
\textbf{SP-STM on the surface of Fe$_{1+y}$Te bulk and thin films.} 
In spin-polarized scanning tunneling microscopy the tunneling current depends on the relative orientation of the tip magnetization and the local spin direction of the
sample. The total tunneling current $I_{\text{t}}$ can be described
by $I_{\text{t}} = I_0 + I_{\text{p}}$, where $I_0$ is the spin-averaged tunneling current and $I_{\text{p}}$ is the spin-polarized tunneling current which is proportional to the product of the spin-polarization $P_{\text{t}}$ of tip and $P_{\text{s}}$ of
the sample, $I_{\text{p}} \propto P_{\text{t}}\cdot P_{\text{s}} \cdot \cos(\beta)$. Here, $\beta$ is the angle between the
spin direction of the tip and the sample.

In general, Fe-coated W-tips have a magnetization direction perpendicular to the tip axis and thus parallel to the surface of the sample.
Therefore, Fe-coated W-tips exhibit sensitivity to the in-plane component of the sample magnetization. By applying an external magnetic field of about 1~T, the magnetization direction of the tip will be reoriented to the applied field direction~\cite{PhysRevLett.103.157201}. Magnetic fields on the order of a Tesla are orders of magnitude too weak to break the exchange interactions between the Fe atoms in the bicollinear antiferromagnetic structure of Fe$_{1+y}$Te. Therefore, the application of external magnetic fields in different directions enables us to image the different spin components of the bicollinear antiferromagnetic structure by recording constant-current SP-STM images. 
 
\begin{figure*}[ht!]
	\begin{center}
		\includegraphics[width=0.95\columnwidth]{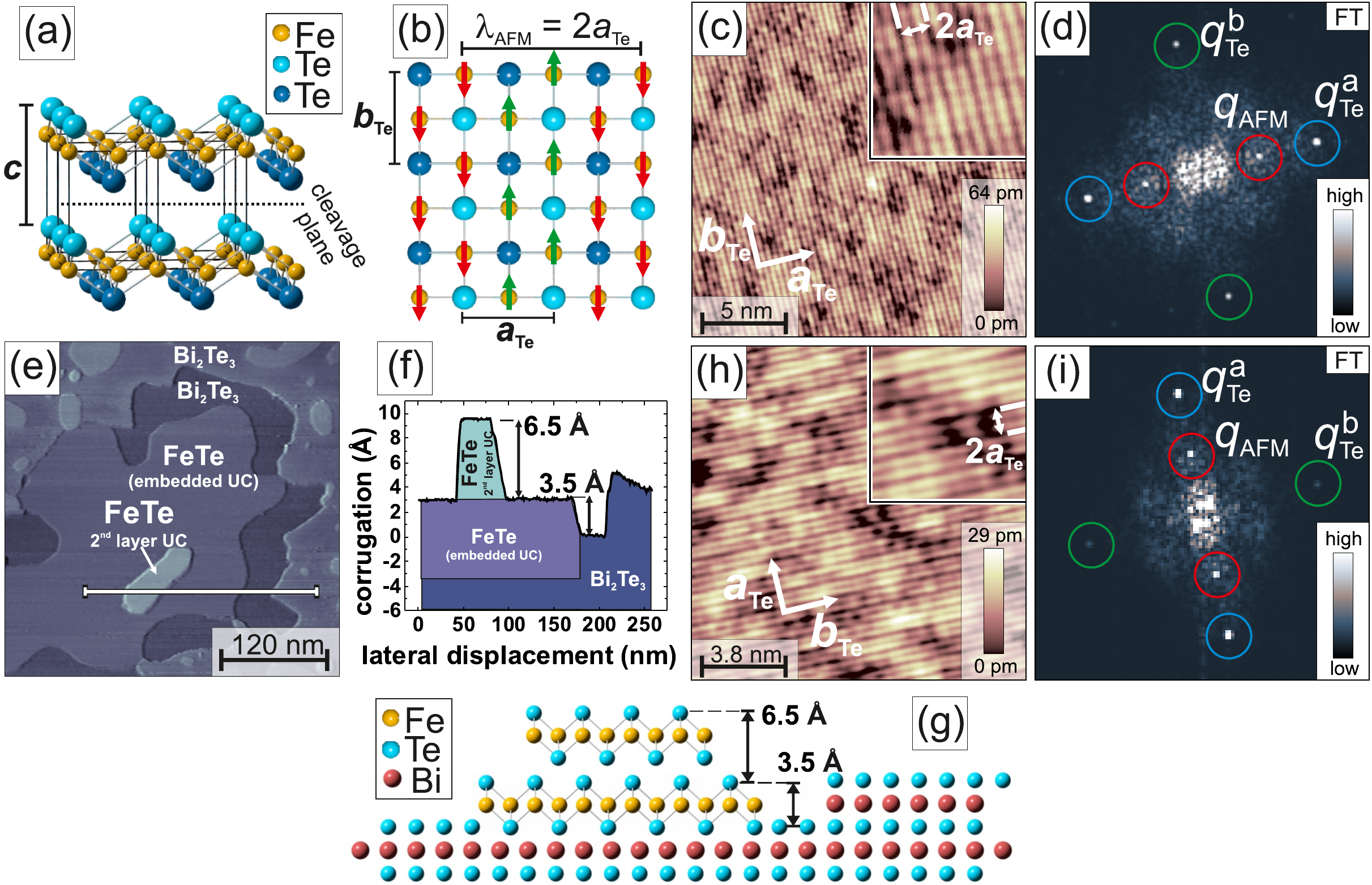}
		\caption{\label{fig:FeTe_simple_topo}\textbf{Structure, morphology and magnetic contrast of the investigated Fe$_{1+y}$Te bulk and thin film samples.} (a) Crystal structure of bulk Fe$_{1+y}$Te showing four unit cells. (b) Top view of the Te terminated surface and the underlying Fe lattice. The spin direction of the bulk bicollinear AFM order is indicated by the red and green arrows (the image is adapted from Ref.~\cite{PhysRevB.79.054503,PhysRevLett.102.247001}). (c) SP-STM image of the surface of Fe$_{1+y}$Te measured with an Fe coated W-tip showing the atomic and the spin structure ($V_{\text{bias}}=+50$~mV, $I_{\text{t}}=320$~pA, $B=0$~T). White arrows denote
		the lattice directions $a_{\text{Te}}$ and $b_{\text{Te}}$. (Inset) Magnified image showing the atomic lattice of the surface Te lattice with a $2a_{\text{Te}}$ periodic superstructure. (d) shows the FT of (c) indicating the Bragg peaks $q^a_{\text{Te}}$ (blue circles), the Bragg peaks $q^b_{\text{Te}}$ (green circles) and the peaks $q_{\text{AFM}}$ of the $2a_{\text{Te}}$ magnetic superstructure (red circles). (e) Topographic overview of the second layer island growth of Fe$_{1+y}$Te on Bi$_2$Te$_3$. (f) Plotted line section along the white line shown in (e). (g) Model of the investigated morphology of Fe$_{1+y}$Te grown on Bi$_2$Te$_3$ along the section in (f). (h) SP-STM image of the two UC thin layer of Fe$_{1+y}$Te grown on Bi$_2$Te$_3$ measured with an Fe coated W-tip showing the atomic structure together with an additional $2a_{\text{Te}}$ spin contrast ($V_{\text{bias}}=+33$~mV, $I_{\text{t}}=4.1$~nA, $B=2.5$~T out-of-plane). White arrows denote the lattice directions $a_{\text{Te}}$ and $b_{\text{Te}}$. (Inset) Magnified image showing the atomic lattice of the surface Te lattice with a $2a_{\text{Te}}$ periodic superstructure. (i) shows the FT of (h) with the same assignment of the different peaks as in (d).}
	\end{center}
\end{figure*}

In Fig.~\ref{fig:FeTe_simple_topo} we give an introduction to both Fe$_{1+y}$Te samples used in this work. The first investigated sample shown in Figs.~\ref{fig:FeTe_simple_topo}(c,d) is the surface of cleaved bulk Fe$_{1+y}$Te ($y\sim 0.07$). Fig.~\ref{fig:FeTe_simple_topo}(c) displays a typical spin-resolved constant-current image of the bulk Fe$_{1+y}$Te surface which shows clear atomic resolution of the Te terminated surface (a detailed analysis of the investigated spin contrast is discussed later on). Compared to previously reported results~\cite{machida1,Sugimoto201385,Enayat08082014} a large area of the Fe$_{1+y}$Te surface is free of excess Fe atoms which is due to the annealing procedure as described in the methods section. This annealing procedure leads to the formation of Fe clusters containing all the excess Fe (outside of the field of view of Fig.~\ref{fig:FeTe_simple_topo}(c)) and large areas with no surface excess Fe atoms in between these clusters. The surface is atomically flat but shows a small variation in the apparent height which is probably caused by excess Fe atoms between the sub-surface layers. 
By calculating the Fourier transform (FT) of Fig.~\ref{fig:FeTe_simple_topo}(c), the lattice periodicity is displayed as bright Bragg spots labeled as $q^a_{\text{Te}}$ and $q^b_{\text{Te}}$ in Fig.~\ref{fig:FeTe_simple_topo}(d), were $q^a_{\text{Te}}$ has a higher intensity than $q^b_{\text{Te}}$ (c.f.\ Refs.\cite{machida1,PhysRevB.93.041101,Enayat08082014}). In addition to the atomic periodicity the constant-current image in Fig.~\ref{fig:FeTe_simple_topo}(c) shows the typical superstructure which has a periodicity of $\lambda=2a_{\text{Te}}$ along the $a_{\text{Te}}$-direction. The presence of this superstructure is visible as additional spots in the FT (labeled with $q_{\text{AFM}}$) with a wave vector of $q_{\text{AFM}}$ = $\frac{1}{2}q^a_{\text{Te}}$. Overall the cleaved surface shows all the characteristics previously reported for Fe$_{1+y}$Te (cf.~Refs.~\cite{PhysRevB.90.224503,Enayat08082014,PhysRevB.93.041101}).

The second investigated sample is a thin Fe$_{1+y}$Te film grown on Bi$_2$Te$_3$ (Figs.~\ref{fig:FeTe_simple_topo}(e-g)). A large scale STM topography of the sample area is displayed in Fig.~\ref{fig:FeTe_simple_topo}(e) together with the height profile along a line shown in Fig.~\ref{fig:FeTe_simple_topo}(f). From atomically resolved images taken on the different visible layers and their apparent heights, we deduce the structure of the layers as sketched in Fig.~\ref{fig:FeTe_simple_topo}(g). The first UC thin layer of FeTe which has an apparent hight of $\sim$3.5~{\AA} (Fig.~\ref{fig:FeTe_simple_topo}(e)) is embedded into an incomplete Bi$_2$Te$_3$ quintuple layer. Due to the interaction with the underlying substrate it exhibits a network of stripe-like dislocations leading to a rough surface (not resolved in the large scale image of Fig.~\ref{fig:FeTe_simple_topo}(e)). On top of the embedded layer the growth of an additional FeTe layer has started and forms a second layer island with a height of ~6.5~{\AA} which is roughly equal
to the $c$-axis lattice constant (6.26~{\AA})~\cite{PhysRevB.79.054503}) of the bulk Fe$_{1+y}$Te UC as shown in Fig.~\ref{fig:FeTe_simple_topo}(a). This second layer has an atomically flat surface.  
In this work we focused on SP-STM measurements on top of the second layer islands indicated by the white arrow in Fig.~\ref{fig:FeTe_simple_topo}(e). In Fig.~\ref{fig:FeTe_simple_topo}(h) an atomically resolved image of the surface on the second layer Fe$_{1+y}$Te island is shown. We do not observe excess Fe atoms on the surface, but we cannot exclude that there is a small amount of excess Fe atoms sitting in the van der Waals gap in between the two FeTe layers. In addition to the atomic corrugation the characteristic $2a_{\text{Te}}$ superstructure is visible as indicated in the inset of Fig.~\ref{fig:FeTe_simple_topo}(h), which is very similar to the one at the bulk Fe$_{1+y}$Te surface shown in Fig.~\ref{fig:FeTe_simple_topo}(c).  This is also visible in the FT displayed in Fig.~\ref{fig:FeTe_simple_topo}(i), which coincides with the FT pattern in Fig.~\ref{fig:FeTe_simple_topo}(d); the atomic lattice shows peaks of similar intensity (labeled with $q^a_{\text{Te}}$ and $q^b_{\text{Te}}$) and also the $2a_{\text{Te}}$ superstructure peaks show similar intensities (labeled with $q_{\text{AFM}}$). Therefore, the spin-resolved STM image of the second layer Fe$_{1+y}$Te
 island exhibits a very similar spin structure as the surface of bulk Fe$_{1+y}$Te.

\begin{figure*}[ht!]
	\begin{center}
		\includegraphics[width=0.8\columnwidth]{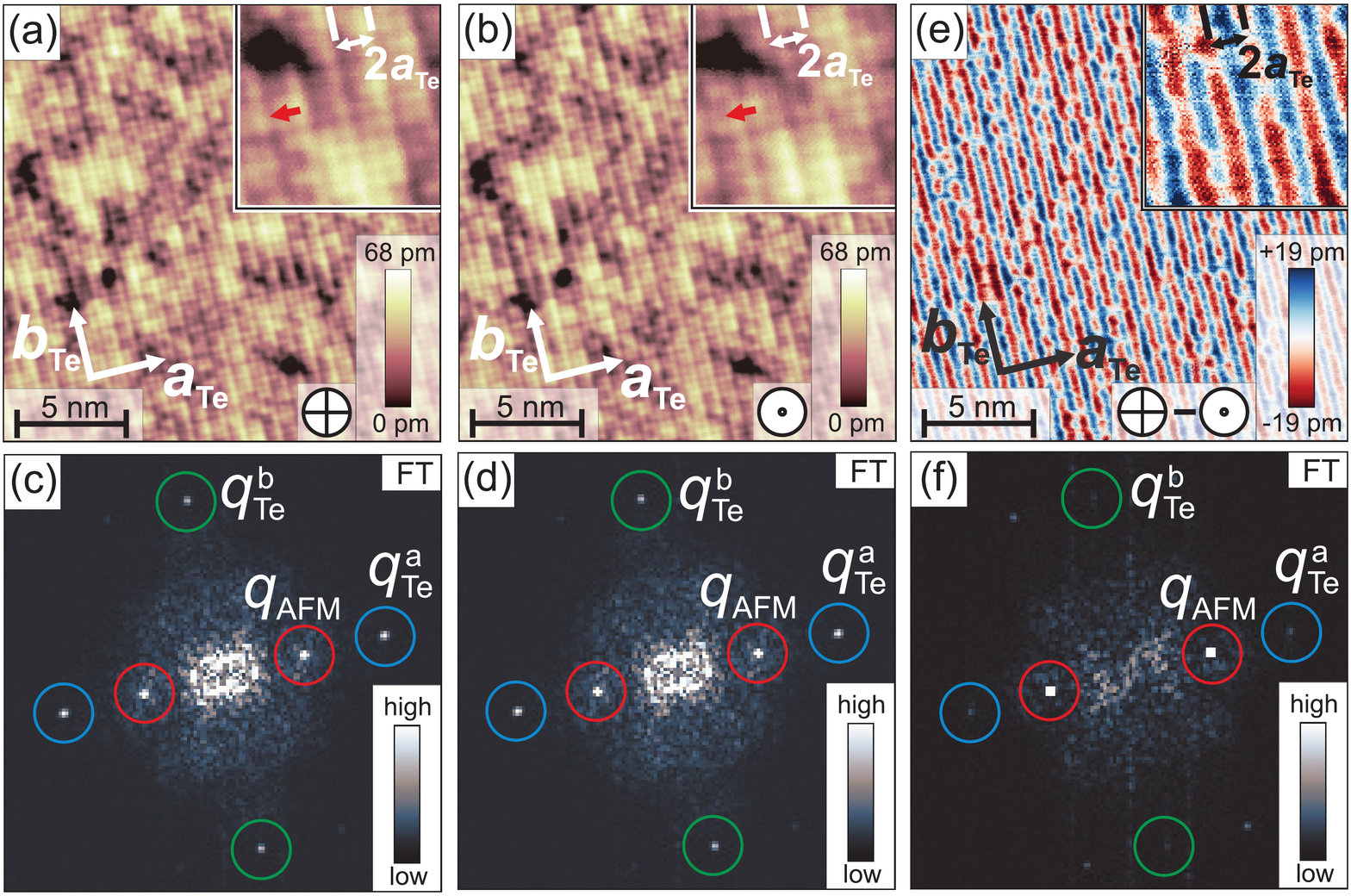}
		\caption{\label{fig:magnetic_contrast}\textbf{SP-STM images revealing the bicollinear antiferromagnetism
			at the surface of bulk Fe$_{1+y}$Te.} For the constant-current images of (a,b) ($V_{\text{bias}}=+50$~mV and $I_{\text{t}}=340$~pA) an out-of-plane external magnetic field of $B=\pm$1~T was applied. The direction of the external magnetic field is indicated by the arrows pointing into or out of the surface plane. White arrows denote the lattice directions $a_{\text{Te}}$ and $b_{\text{Te}}$. (Insets) Magnified images showing the atomic lattice of the chalcogen terminated Te lattice with a $2a_{\text{Te}}$ periodic superstructure. The red arrow denotes the same location on the sample, where due to the opposite direction of the tip magnetization in (a) a minimum of the  $2a_{\text{Te}}$ modulation is observed and in (b) a maximum. (c,d) show the FTs of the constant-current maps in (a,b). (e)
						Difference image of (a,b) consistent with a bicollinear antiferromagnetic
						structure of the Fe$_{1+y}$Te surface. (Inset) Magnified image showing the $2a_{\text{Te}}$ periodic superstructure. (f) displays the corresponding FT of (e). 
			 The peaks in the FTs (c), (d) and (f) are labeled with blue circles ($q^a_{\text{Te}}$), green circles ($q^b_{\text{Te}}$) and red circles ($q_{\text{AFM}}$).}
	\end{center}
\end{figure*}

\textbf{Magnetic field dependent SP-STM on the surface of bulk Fe$_{1+y}$Te.}
After introducing the two different sample systems used in this work we will now discuss on our results obtained with different orientations of the tip-magnetization starting with the surface of bulk Fe$_{1+y}$Te and out-of plane tip sensitivity. Figures~\ref{fig:magnetic_contrast}(a,b) show constant-current maps of the same surface area obtained with an Fe-coated tip. They were recorded in a magnetic field of 1~T with opposite out-of-plane field directions which forces the tip magnetization to point up and down. The two SP-STM images show the characteristic $2a_{\text{Te}}$ superstructure which is also visible by the $q_{\text{AFM}}$ peak displayed in the corresponding FTs in Fig.~\ref{fig:magnetic_contrast}(c,d). In contrast to different interpretations such as
charge ordering phenomena~\cite{Sugimoto201385,fete_co}, this additional superstructure
has been attributed to a direct imaging of the
bicollinear antiferromagentic order of Fe$_{1+y}$Te by
SP-STM~\cite{Enayat08082014,udaj}. Our experiments show that this is indeed the case and confirm that spin-polarized tunneling is the origin of the additional $2a_{\text{Te}}$ superstructure by applying external magnetic
fields. By comparing the superstructure with the atomic lattice, a phase shift by one lattice unit is observed for opposite magnetic field directions (cf.~insets of Figs.~\ref{fig:magnetic_contrast}(a,b)). By subtracting the constant-current maps in Fig.~\ref{fig:magnetic_contrast}(a) from that in Fig.~\ref{fig:magnetic_contrast}(b) an image of the out-of plane components of the spin structure is obtained in Fig.~\ref{fig:magnetic_contrast}(e) which mainly shows a stripe pattern with a periodicity of $2a_{\text{Te}}$ along the $a_{\text{Te}}$-direction. This is also reflected by the FT of the difference image in Fig.~\ref{fig:magnetic_contrast}(f). Here only the peak $q_{\text{AFM}}$ remains and the Bragg peaks $q^a_{\text{Te}}$ and $q^b_{\text{Te}}$ related to the atomic lattice have vanished. This observation directly confirms the results from Ref.~\cite{Enayat08082014} and proves spin-polarized tunneling contrast due to the bicollinear antiferromagnetic spin structure of Fe$_{1+y}$Te. The maximum spin contrast appears between every second Fe lattice site located between two neighboring Te sites. This
can be explained by the fact that spin-polarized tunneling primarily results from the 3d states of
Fe being located below the top Te layer.

However, from the strong spin contrast we see using the out-of plane sensitive magnetic tip, we can additionally conclude, that the surface antiferromagnetic structure has a considerable out-of plane spin component. Considering the magnetic structure of bulk Fe$_{1+y}$Te as known from neutron diffraction\cite{Fuchart,PhysRevB.79.054503,PhysRevLett.102.247001} depicted in Fig.~\ref{fig:FeTe_simple_topo}(b) this leads to the conclusion, that the surface spins are reoriented with respect to the corresponding bulk layers.

\begin{figure*}[ht!]
	\begin{center}
		\includegraphics[width=0.98\columnwidth]{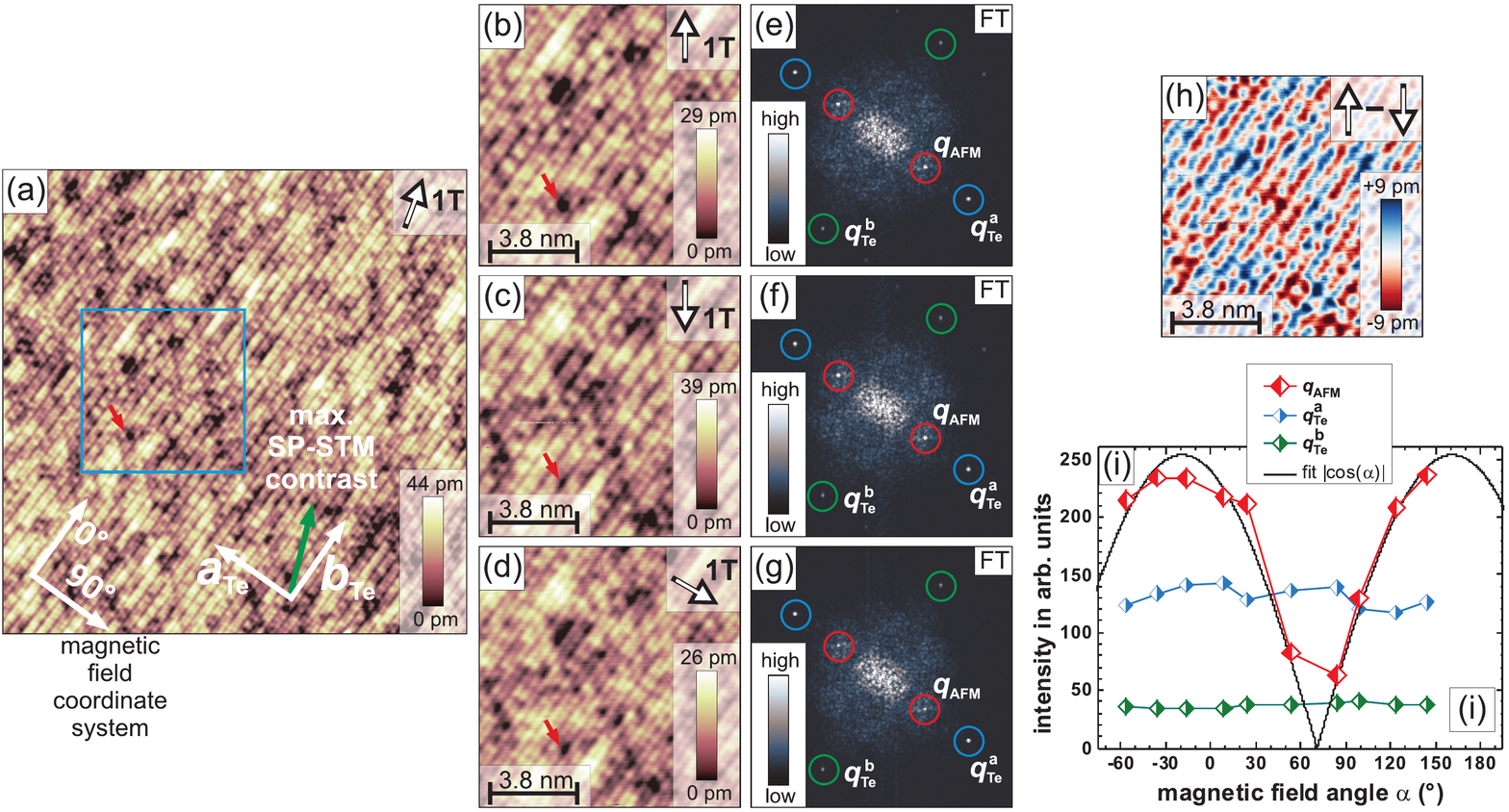}
		\caption{\label{fig:in-plane_neu}\textbf{In-plane tip-magnetization direction dependent spin contrast at the surface of bulk Fe$_{1+y}$Te}. (a) Spin-resolved ($32.5 \times 32.5$)~nm$^2$ overview measured in an in-plane external magnetic field of 1~T at -16$^\circ$ ($V_{\text{bias}}=+50$~mV and $I_{\text{t}}=500$~pA). 
			(b) and (d) show the magnetic contrast at opposite field directions at -36$^\circ$ and 144$^\circ$ ($|B|=1$~T) revealing a phase shift, which is shown in the difference image in (h). (d) shows almost vanishing magnetic contrast at an angle of 84$^\circ$ ($|B|=1$~T). The red arrows in (b-d) indicate an atomic scale defect, i.e.\ point at the identical positions in all images. In all SP-STM images the direction of the applied field is indicated by the arrows in the insets. (e-g) show the FT of (b-d), respectively. The spots in the FTs are labeled with blue circles ($q^a_{\text{Te}}$), green circles ($q^b_{\text{Te}}$) and red circles ($q_{\text{AFM}}$). (i) shows the plotted intensity of the three components $q^a_{\text{Te}}$, $q^b_{\text{Te}}$ and $q_{\text{AFM}}$ of the FTs. The color coding of the corresponding symbols match the circular markings in the FTs.}
	\end{center}
\end{figure*}

In order to analyze the in-plane surface spin components of the bicollinear antiferromagnetic structure of Fe$_{1+y}$Te, we performed experiments within a vector-magnet system which enables to rotate the tip magnetization direction within the film plane. For this purpose a magnetic field with a fixed amplitude was applied parallel to the surface and then stepwise rotated for each taken SP-STM image recorded. The results are shown in Fig.~\ref{fig:in-plane_neu} for the magnetic field amplitude of 1~T. Figure~\ref{fig:in-plane_neu}(a) shows an overview of the atomically resolved Fe$_{1+y}$Te surface with a magnetic field applied under an angle of $\alpha = -16^\circ$ relative to the given magnetic field coordinate system. On top of the atomic corrugation the strong $2a_{\text{Te}}$ superstructure is visible running continuously through the whole image from the bottom left to the top right. The blue square indicates the area where the investigations with rotated magnetic fields were performed and the red arrow indicates a defect used as a marker for the atomic-scale registry. Upon rotating the magnetic field parallel to the surface the intensity of the $2a_{\text{Te}}$ superstructure was measured by taking spin-resolved constant-current maps for each field direction (Figs.~\ref{fig:in-plane_neu}(b-d), see full set of images in Supplementary Fig.~S1). The intensity of the superstructure was then extracted from the amplitude of the $q_{\text{AFM}}$-peak in the FTs (Figs.~\ref{fig:in-plane_neu}(e-g)). In addition to the $q_{\text{AFM}}$-peak the intensity of the Bragg peaks $q^a_{\text{Te}}$ $q^b_{\text{Te}}$ were recorded as a reference in order to confirm, that the tip did not change during the full magnetic field sweep.  These intensities are shown in Fig.~\ref{fig:in-plane_neu}(i) as a function of the magnetic field angle. In Figs.~\ref{fig:in-plane_neu}(c,e), the tip magnetization direction is pointing in opposite directions within the surface plane. It is again apparent that the $2a_{\text{Te}}$ superstructure observed for both field directions exhibits a phase shift of one lattice unit. This is verified by calculating the difference of Figs.~\ref{fig:in-plane_neu}(b,c) shown in (h), where only the magnetic signal of the $2a_{\text{Te}}$ superstructure remains. By comparing the intensities of the two Bragg peaks $q^a_{\text{Te}}$ and $q^b_{\text{Te}}$ and  the bicollinear $q_{\text{AFM}}$ peak extracted from the FTs of these images shown in Figs.~\ref{fig:in-plane_neu}(e,f) we can conclude, that both SP-STM images exhibit the same amplitudes for the atomic corrugation and the $2a_{\text{Te}}$ superstructure, respectively as shown in the plot of Fig.~\ref{fig:in-plane_neu}(i). In contrast, in the SP-STM image and its FT for a field direction of 84$^\circ$ (Fig.~\ref{fig:in-plane_neu}(d,g) the amplitude of the atomic corrugation remains at the same level but the amplitude of the $2a_{\text{Te}}$ superstructure is strongly reduced. Overall the resulting angular dependence plotted in Fig.~\ref{fig:in-plane_neu}(i) reveals a periodic variation of the $q_{\text{AFM}}$ peak intensity with a  periodicity of 180$^\circ$, while the intensities of the Bragg peaks do not show significant changes. As shown by the fitted line, the $q_{\text{AFM}}$ intensity nicely follows a $|\text{cos}(\alpha)|$-dependence which is expected for spin-polarized tunneling into an antiferromagnetic spin structure upon the rotation of the tip magnetization. The maximum of the intensity was found to be at ~-19$^\circ$. The corresponding in-plane spin direction of the bicollinear antiferromagnetic structure is indicated by the green arrow in Fig.~\ref{fig:in-plane_neu}(a). We can thus conclude, that the spin direction of the surface layer bilcollinear antiferromagnetic structure deviates by 19$^\circ$ from the $b_{\text{Te}}$-direction, which is the spin direction of the bicollinear structure in bulk Fe$_{1+y}$Te.

\begin{figure*}[ht!]
\begin{center}
\includegraphics[width=0.95\columnwidth]{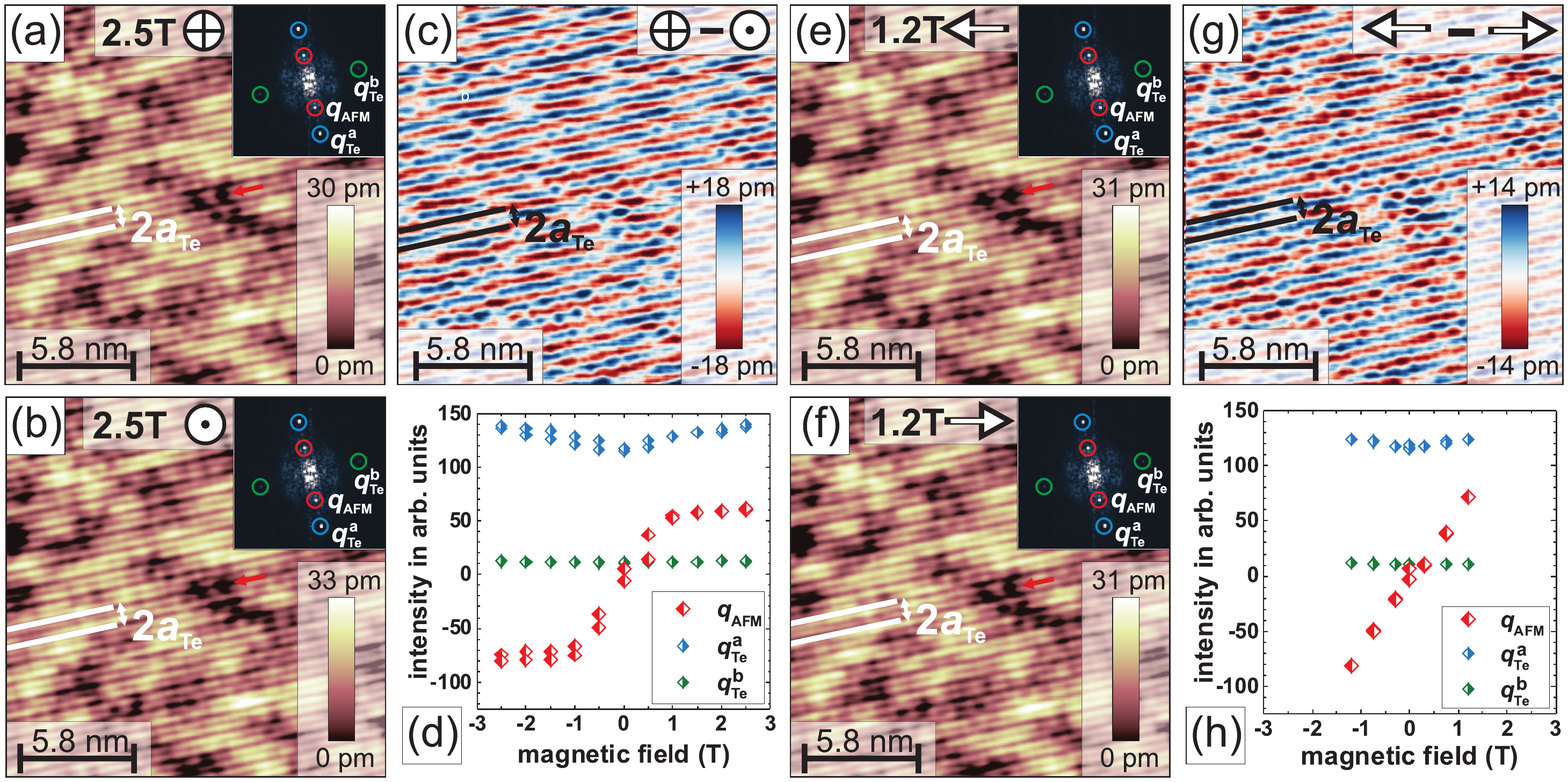}
\caption{\label{fig:mag_contrast_layer_neu}\textbf{SP-STM  revealing bicollinear antiferromagnetism
			at the surface of Fe$_{1+y}$Te thin films grown on Bi$_2$Te$_3$.} (a),(b) and (e),(f) show the magnetic contrast the same area of a on two-layer thick Fe$_{1+y}$Te island measured with the same tip where in (a),(b) $\pm$2.5~T were applied in the out-of-plane direction, whereas in (e),(f) $\pm$1.2~T were applied in the in-plane direction ($V_{\text{bias}}=+33$~mV and $I_{\text{t}}=4.1$~nA). The red arrows indicate a defect used as a marker. The direction of the applied magnetic field is indicated by the arrows.  The insets in the upper right of (a),(b) and (e),(f) display the FTs of each SP-STM image and the spots in the FTs are labeled with blue circles ($q^a_{\text{Te}}$), green circles ($q^b_{\text{Te}}$) and red circles ($q_{\text{AFM}}$). (c) shows the difference image of (a,b), whereas (g) displays the difference image of (e,f). The magnetic field dependence of the intensity of $q^a_{\text{Te}}$, $q^b_{\text{Te}}$ and $q_{\text{AFM}}$ in the FTs for out-of-plane magnetic fields is plotted in (d) and for magnetic fields applied in the in-plane direction in (h). The color coding of the corresponding symbols match the circular markings in the FTs. Since the absolute value of the FT contains no phase information the amplitude of the $q_{\text{AFM}}$ peak was multiplied by -1 for negative field directions to illustrate the observed phase shift of the bicollinear superstructure.} 
\end{center}
\end{figure*}

\textbf{Magnetic field dependent SP-STM on the surface of thin Fe$_{1+y}$Te films on Bi$_2$Te$_3$.}
For comparison, we will now discuss the results obtained for a thin Fe$_{1+y}$Te film grown on a Bi$_2$Te$_3$ substrate. All SP-STM data in Fig.~\ref{fig:mag_contrast_layer_neu} were obtained in the same surface area on the top of a UC high Fe$_{1+y}$Te island for magnetic fields of 2.5~T applied in opposite out-of-plane directions (Figs.~\ref{fig:mag_contrast_layer_neu}(a,b)) and for magnetic fields of 1.2~T applied in opposite in-plane directions (Figs.~\ref{fig:mag_contrast_layer_neu}(e,f)). As for the measurements discussed above, the characteristic $2a_{\text{Te}}$ superstructure is observed in all four SP-STM data sets. The FTs in the insets of Figs.~\ref{fig:mag_contrast_layer_neu}(a,b) and (e,f) also show the same $q_{\text{AFM}}$ pattern as for the bulk samples. By comparing the position of the maximum of the $2a_{\text{Te}}$ superstructure relative to the underlying Te-lattice, e.g.\ at the position indicated by the red arrow, the Fe$_{1+y}$Te thin film system also reveals a phase shift by one lattice constant upon inverting of the tip's magnetization direction in the out-of-plane as well as in the in-plane direction. This is also obvious from the difference images in Figs.~\ref{fig:mag_contrast_layer_neu}(c,g) reflecting the images of the out-of plane and in-plane components of the spin structure, respectively. Therefore, we can conclude that the surface of the thin film Fe$_{1+y}$Te on Bi$_2$Te$_3$ exhibits the same bicollinear antiferromagnetic spin structure as the bulk samples. Additionally, in order to analyze the direction of the surface spins, the magnetic field dependence of the intensities of the $q_{\text{AFM}}$ peak and of the Bragg peaks $q^a_{\text{Te}}$ and $q^b_{\text{Te}}$ were extracted for the out-of-plane and for the in-plane direction and are shown in Figs.~\ref{fig:mag_contrast_layer_neu}(d,h), respectively (full series of magnetic field dependent SP-STM images is given in the supplementary Figs.~S2 and S3). 
Upon increasing the magnetic field in the out-of-plane direction the intensity of $q_{\text{AFM}}$ starts changing since the magnetization direction of the tip is continuously rotated from the in-plane to the out-of-plane direction as shown in Fig.~\ref{fig:mag_contrast_layer_neu}(d). For both the negative and the positive field direction the intensity of $q_{\text{AFM}}$ saturates when the tip magnetization is fully rotated into the out-of-plane direction. For the in-plane direction a similar behavior is observed as shown in Fig.~\ref{fig:mag_contrast_layer_neu}(h). Here, in zero magnetic field the initial tip magnetization has an unknown orientation with respect to the spin direction of the Fe atoms in the surface Fe$_{1+y}$Te layer. Upon increasing the magnetic field, the tip continuously rotates into the direction of the magnetic field, where a strong magnetic contrast is deduced by the increase in the $q_{\text{AFM}}$ intensity. From the similar saturation values of the $q_{\text{AFM}}$ intensities in Figs.~\ref{fig:mag_contrast_layer_neu}(d,h) we can conclude, that the out-of plane component of the bicollinear antiferromagnetic spin structure in the surface layer of the Fe$_{1+y}$Te thin film has a similar strength as the in-plane component. 
\section{Discussion\label{Discussion}}
In summary, we provide a direct proof that spin-polarized tunneling is responsible for the observation of the $2a_{\text{Te}}$ superstructure in scanning tunneling microscopy on Fe$_{1+y}$Te by using well defined spin-sensitive Fe-coated W-tips. Measurements under applied magnetic fields reveal that the $2a_{\text{Te}}$ superstructure can only be interpreted in terms of direct SP-STM imaging of the bicollinear AFM order of the Fe$_{1+y}$Te surface. This confirms previous SP-STM results with magnetically sensitive tips on Fe$_{1+y}$Te~\cite{Enayat08082014,udaj} and does not support the interpretation of a $2a_{\text{Te}}$ CDW-order discussed in Ref.~\cite{Sugimoto201385,PhysRevB.90.224503}. 
Furthermore, we have found a strong contribution of the surface spin component in the out-of-plane direction along the c-axis and an in-plane component which deviates from the $b_{\text{Te}}$-axis direction, for both the bulk Fe$_{1+y}$Te samples and the thin Fe$_{1+y}$Te films grown on Bi$_2$Te$_3$. Neutron scattering, which is sensitive to the bulk magnetization, revealed a dominant spin direction of the bicollinear antiferromagnetic structure along the $b_{Te}$-axis direction~\cite{PhysRevB.79.054503,PhysRevLett.102.247001,PhysRevB.84.064403}. A tiny component of the magnetization along the $a$- and $c$-axes has been mainly attributed to local moments of excess iron atoms which are located in the Van der Waals gap between the FeTe layers~\cite{PhysRevB.79.054503}. In contrast, our experiments indicate a strong component of the surface magnetization out of the $b_{Te}$-axis direction, in favour of a reorientation of the ordered magnetic moments \textit{at the surface} of Fe$_{1+y}$Te. The central question is, thus, what drives this surface reorientation of the spin direction by keeping the overall bicollinear order. Due to its layered crystal structure bulk Fe$_{1+y}$Te has a quasi-two-dimensional electronic structure and the \textit{relative} orientation of the spins within the bicollinear structure is mostly determined by the exchange interaction of the Fe atoms within the $a$-$b$-plane of a given Fe$_{1+y}$Te layer. This exchange interaction is relatively strong and largely unaffected by additional effects occurring at the surface such as lattice relaxations or charge redistribution. Therefore, the bicollinear antiferromagnetic structure is preserved at the surface of Fe$_{1+y}$Te. On the other hand, the \textit{absolute} orientation of the magnetic moments is influenced by the magnetic anisotropy energy  which has a much smaller energy scale. Typically, the magnetic anisotropy is strongly influenced by electronic effects or lattice relaxations. Based on these considerations, we propose that the observed reorientation of the spin structure at the surface of Fe$_{1+y}$Te is induced by such surface effects. Our findings might have implications for related sample systems. E.g.\ comparable spin-polarized STM studies could offer insight into the origin of the strong increased $T_{\text{C}}$ of the monolayer FeSe grown on SrTiO$_3$ compared to bulk FeSe~\cite{0256-307X-29-3-037402,FeUCARPES,nmat4153}. Although bulk FeSe does not show static magnetic order, spin fluctuations are known to exist and being coupled to superconducting pairing~\cite{RevModPhys.87.855}, which likewise might be strongly modified on the surface. 

In conclusion, we have shown that SP-STM experiments with Fe-coated W-tips are well suited to investigate correlated electron systems such as Fe$_{1+y}$Te with a complex electronic and magnetic structure. However, a complete characterization requires SP-STM experiments performed in 3D-vector-field systems offering field-dependent studies with arbitrary field orientation. 
\section{Methods\label{methods}}
\textbf{Tip and sample preparation, experimental techniques.} High quality Fe$_{1+y}$Te single crystals were synthesized
using the flux method~\cite{PhysRevB.80.174509} where the excess Fe ratio $y$ was kept as low as possible. The measured composition of the crystals using single crystal x-ray diffraction resulted in $y \approx 7\%$.
The samples were cleaved \textit{in situ} at room temperature (RT), and measured
in ultra-high vacuum (UHV) with a background pressure better than $3\times10^{-10}$~mbar. For all SP-STM measurements the bulk samples were moderately annealed at 100~$^{\circ}$C after cleaving which removes the surface excess iron and leads to an atomically flat surface. Additionally, ultra-thin
Fe$_{1+y}$Te films were grown \textit{in situ} on Bi$_2$Te$_3$ substrates. Single crystals of Bi$_2$Te$_3$ were synthesized using a Stockbarger method and were well characterized using ARPES~\cite{2016arXiv160309689V}. Fe-chalcogenide thin film preparation was carried out in a UHV system with a base pressure better than $3\times10^{-10}$~mbar. The Bi$_2$Te$_3$ crystals were cleaved \textit{in-situ} under UHV conditions and Fe$_{1+y}$Te thin films were prepared by depositing 0.5-1 monolayer (ML) Fe onto a clean Bi$_2$Te$_3$ surface at RT followed by a 15~min annealing cycle at $\sim$300~$^{\circ}$C. Fe deposited on Bi$_2$Te$_3$ reacts with the substrate upon annealing, most likely via a substitutional process replacing Bi by Fe. This preparation was performed similar to the method described in Ref.~\cite{Cavallin201672}. Here, the main difference in the growth is that the Fe$_{1+y}$Te islands do not show a moir\'e-pattern but a smooth atomically flat surface.

The scanning tunneling microscopy experiments were performed in two home-built low temperature UHV-STM systems with out-of-plane magnetic fields up to 5~T and in-plane magnetic fields up to 2~T~\cite{wittneven,meckler} at the base temperature of 6.5 K. To prepare spin-sensitive tips, electrochemically etched W-tips were shortly heated to $\sim$2000~$^{\circ}$C (flash) and afterwards a thin Fe film of $\sim$10~nm thickness was deposited onto the tip apex by e-beam deposition~\cite{0034-4885-66-4-203,RevModPhys.81.1495}. All STM data were recorded in constant-current mode with a fixed bias voltage $V_{\text{bias}}$ and a constant set-point for the tunneling current $I_{\text{t}}$. The Fourier transforms (FTs) were calculated from the absolute value of the complex  two-dimensional fast Fourier transform which is proportional to the power spectral density.
\section{Acknowledgements\label{acknowledgements}}
We are indebted to Alexander Balatsky, Christopher Triola and Tim O. Wehling for valuable discussions. 
T.~H.\ acknowledges funding from Project No.~HA 6037/2-1 of the DFG. Work at Tulane is supported by the US DOE under grant DE-SC0014208 (for material synthesis). U.~R.~S., S.~M., A.~K.\ and R.~W.\ acknowledge funding via the ERC Advanced Grand ASTONISH (No.~338802). L.~C., Ph.~H.\ and J.~W.\ acknowledge funding through the DFG
priority program SPP1666 (grant No. WI 3097/2). We also thank the Aarhus University Research Foundation for supporting the Bi$_2$Te$_3$ bulk crystal growth. M.~B., E.~M.~J.~H.\ and B.~B.~I.\ acknowledges the financial support for Center of Materials Crystallography (CMC), funded by the Danish National Research Foundation (DNRF93). T.~H.\ acknowledges Maciej Bazarnik and Lorenz Schmidt for support in the STM labs.
\section{Author contributions\label{contributions}}
T.~H.\ and J.~W.\ designed the experiment. T.~H.\ performed the SP-STM experiments on bulk Fe$_{1+y}$Te. T.~H.\ and U.~R.~S.\ have grown and characterized the thin film samples and performed the SP-STM experiments. T.~H., U.~R.~S., S.~M., A.~K.\ and L.~C.\ analyzed the data. T.~H., J.~W.\ and R.~W.\ wrote the manuscript. J.~H.\ and Z.~M.\ have grown the Fe$_{1+y}$Te single crystals. E.~M.J.~H., M.~B., B.~B.~I.\ and Ph.~H.\ have grown and characterized the Bi$_2$Te$_3$ samples and measured the composition of the Fe$_{1+y}$Te single crystals.
\section{Competing financial interests\label{Competing}}
The authors declare no competing financial interests.

%


\afterpage{\includepdf[pages={1-1}]{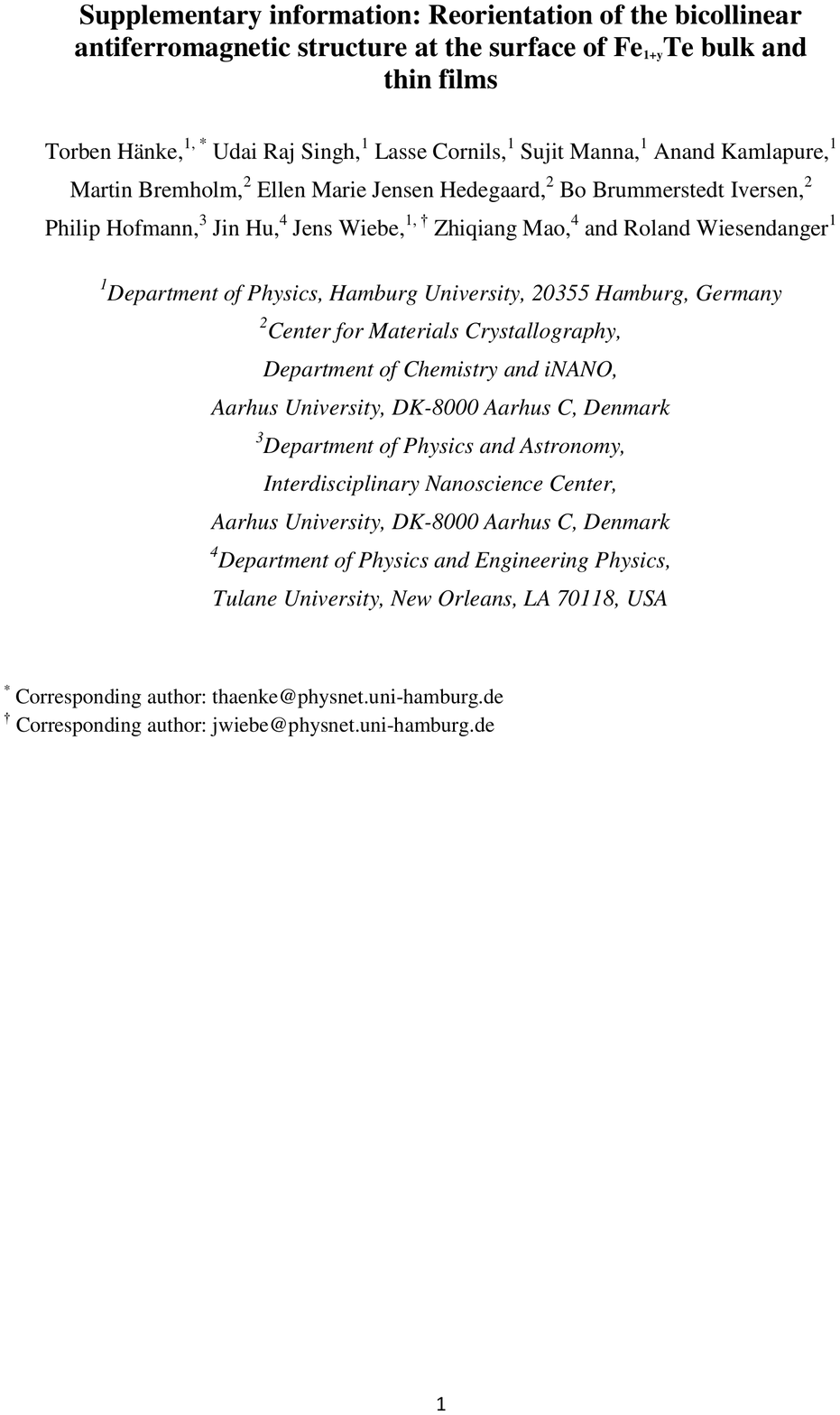}}
\afterpage{\includepdf[pages={2-2}]{fete_supplementary_information.pdf}}
\afterpage{\includepdf[pages={3-3}]{fete_supplementary_information.pdf}}
\afterpage{

\section*{Supplementary material}

# Supplementary information: Reorientation of the bicollinear antiferromagnetic structure at the surface of $Fe_{1+y}$Te bulk and thin films

Torben Hänke,[1,*] Udai Raj Singh,[1] Lasse Cornils,[1] Sujit Manna,[1] Anand Kamlapure,[1] Martin Bremholm,[2] Ellen Marie Jensen Hedegaard,[2] Bo Brummerstedt Iversen,[2] Philip Hofmann,[3] Jin Hu,[4] Jens Wiebe,[1,†] Zhiqiang Mao,[4] and Roland Wiesendanger[1]

[1]*Department of Physics, Hamburg University, 20355 Hamburg, Germany*
[2]*Center for Materials Crystallography,*
*Department of Chemistry and iNANO,*
*Aarhus University, DK-8000 Aarhus C, Denmark*
[3]*Department of Physics and Astronomy,*
*Interdisciplinary Nanoscience Center,*
*Aarhus University, DK-8000 Aarhus C, Denmark*
[4]*Department of Physics and Engineering Physics,*
*Tulane University, New Orleans, LA 70118, USA*

[*] Corresponding author: thaenke@physnet.uni-hamburg.de
[†] Corresponding author: jwiebe@physnet.uni-hamburg.de

**Supplementary Figures:**

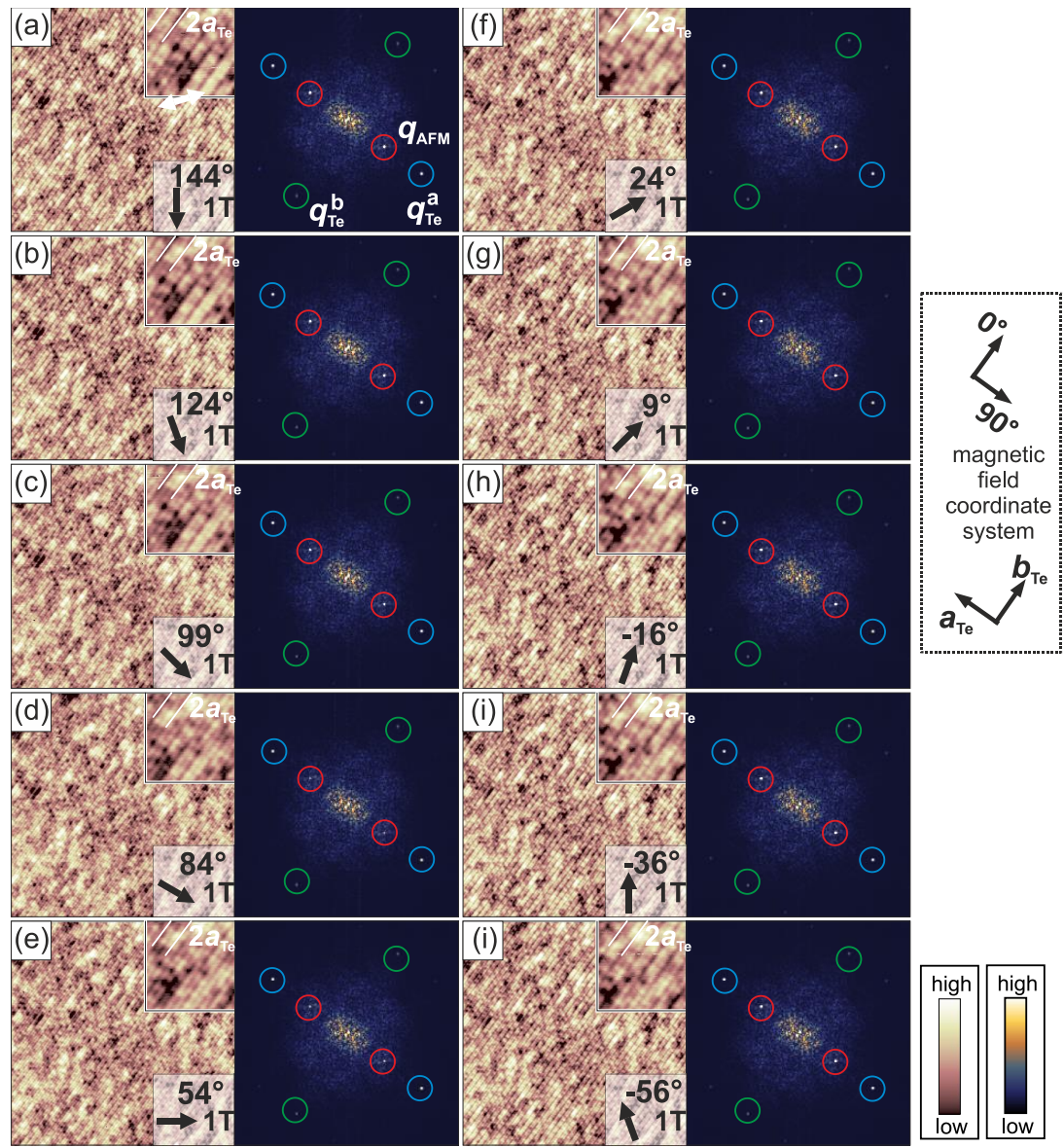

**Fig. S1: Full data set of the in-plane tip-magnetization dependent spin contrast at the surface of bulk $Fe_{1+y}Te$ used for Fig.3(i) of the main manuscript.** (a)-(i) show the spin-resolved constant-current images ($V_{bias}$ = +50 mV, $I_t$ = 320 pA, (32.5×32.5) nm$^2$) of the same field of view for external magnetic fields of 1 Tesla applied in different in-plane directions (left panels) and the corresponding Fourier transforms (FT) (right panels). The direction of the external magnetic field is indicated by the black arrows. The peaks $q_{Te}^a$, $q_{Te}^b$ and $q_{AFM}$ in the FTs are marked with blue, green and red circles, respectively. (inset) Magnified image showing the atomic lattice of the Te-terminated surface and the $2a_{Te}$ periodic superstructure.

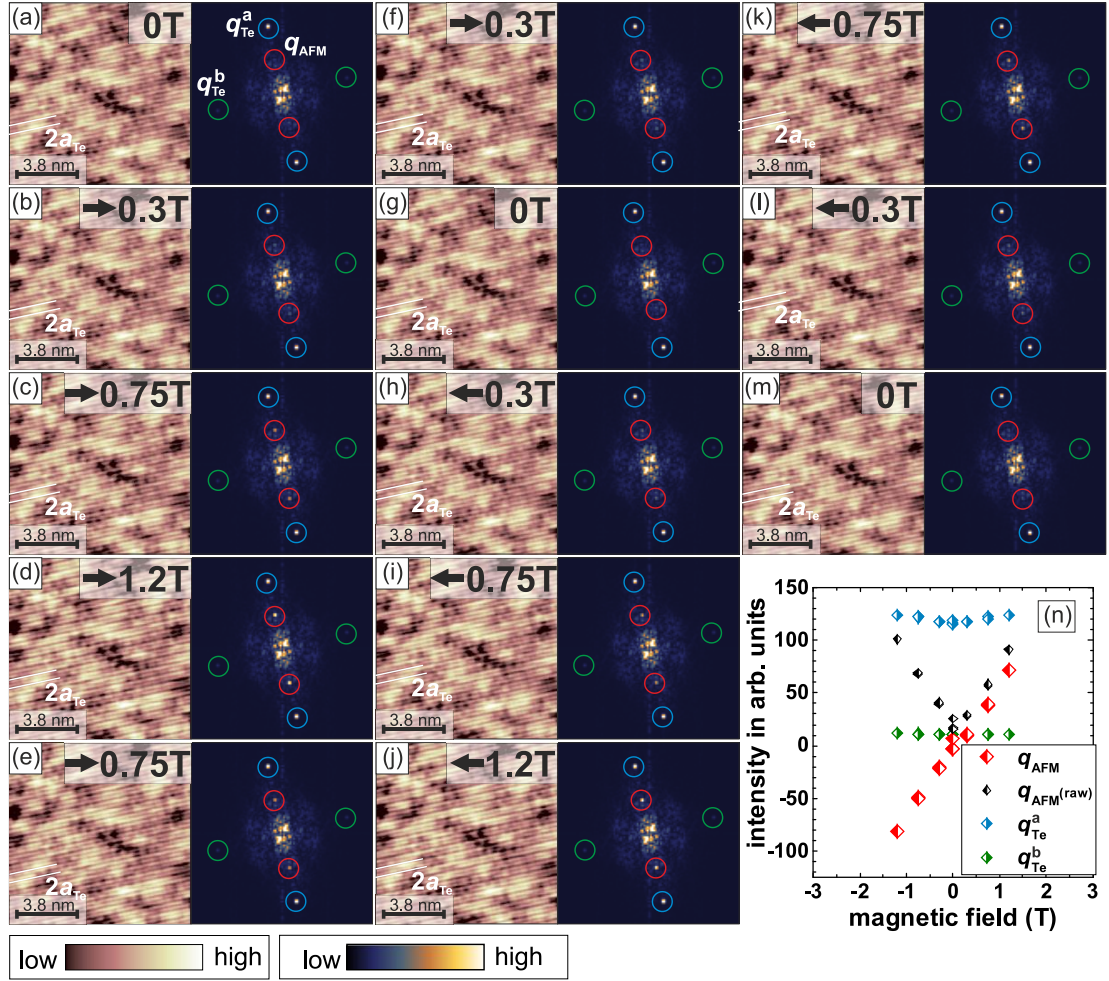

**Fig. S2: Full data set of the in-plane tip-magnetization dependent spin contrast at the surface of thin $Fe_{1+y}Te$ films grown on $Bi_2Te_3$ used for Fig.4(h) of the main manuscript.** (a)-(m) show the spin-resolved constant-current images ($V_{bias}$ = +33 mV, $I_t$ = 4.1 nA) of the same field of view for external magnetic fields of different amplitudes applied in the in-plane direction (left panels) and the corresponding Fourier transforms (FT) (right panels). The direction and the amplitude of the external magnetic field are indicated by the black arrows and the values in the insets. The peaks $q_{Te}^a$, $q_{Te}^b$ and $q_{AFM}$ in the FTs are marked with blue, green and red circles, respectively. The magnetic field dependence of the intensities of $q_{Te}^a$, $q_{Te}^b$ and $q_{AFM}$ in the FTs is plotted in (n), where the color coding of the corresponding symbols match the circular markings in the FTs. For $q_{AFM}$ the extracted raw intensities are plotted with black symbols. It is visible, that the intensity of $q_{AFM}$ is getting larger for both field directions upon increasing the magnetic field amplitude. Since a phase shift of the $2a_{Te}$ superstructure by one lattice constant is observed for opposite field directions and the plotted absolute value of the FT contains no phase information, the amplitude for negative field directions was multiplied by -1 in order to transform the FT intensities into values of relative orientation of the spin-directions (green symbols). Furthermore, the background intensity at $q_{AFM}$ at 0 T was subtracted for all data points of $q_{AFM}$.

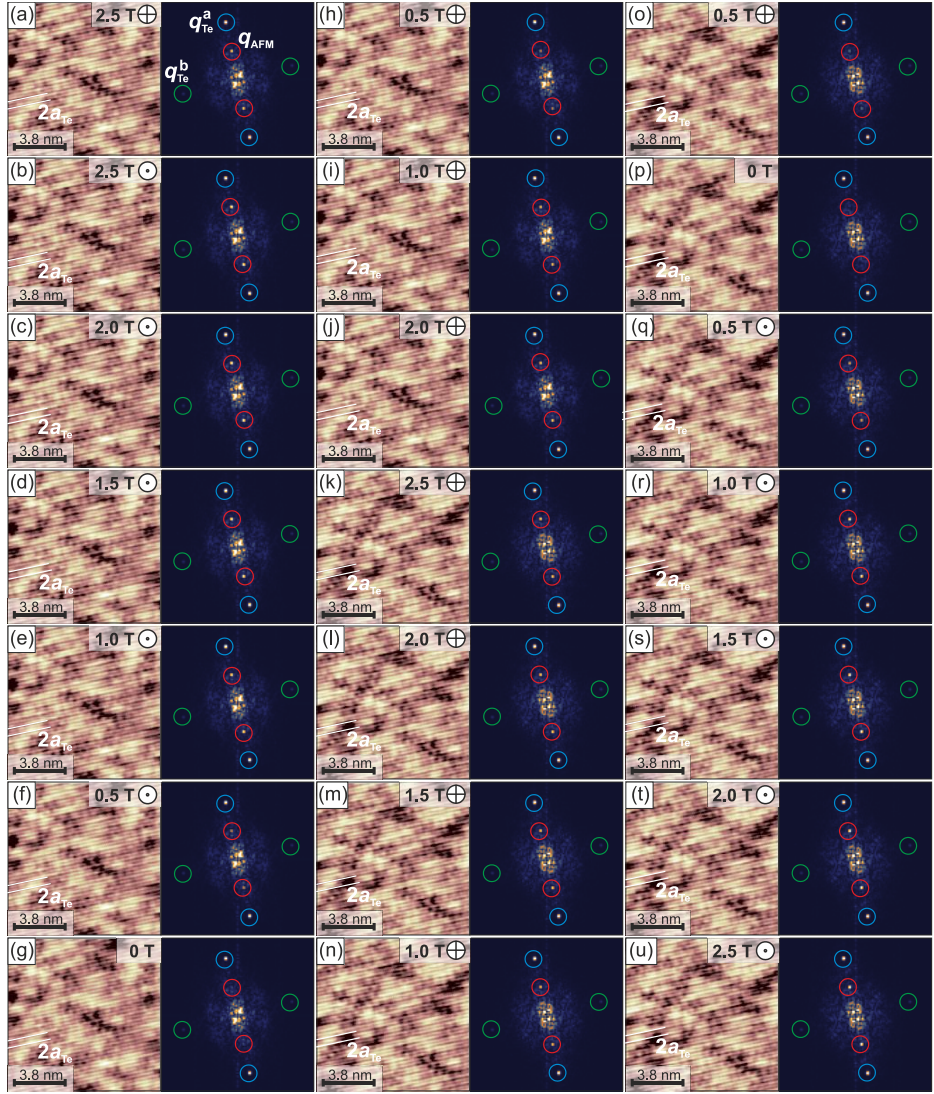

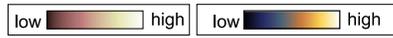

**Fig. S3: Full data set of the out-of-plane tip-magnetization dependent spin contrast at the surface of thin $Fe_{1+y}Te$ films grown on $Bi_2Te_3$ used for Fig. 4(d) of the main manuscript.** (a)-(u) show the spin-resolved constant-current images ($V_{bias}$ = +33 mV, $I_t$ = 4.1 nA) on the same sample area for external magnetic fields of different amplitudes applied in the out-of-plane direction (left panels) and the corresponding Fourier transforms (FT) (right panels). The direction and the amplitude of the external magnetic field are indicated by the black arrow and the value in the insets. The peaks $q_{Te}^a$, $q_{Te}^b$ and $q_{AFM}$ in the FTs are marked with blue, green and red circles, respectively. The magnetic field dependence of the intensities of $q_{Te}^b$, $q_{Te}^b$ and $q_{AFM}$ in the FTs is plotted in (v), where the color coding of the corresponding symbols match the circular markings in the FTs. For $q_{AFM}$ the extracted raw intensities are plotted with black symbols. It is visible, that the intensity of $q_{AFM}$ is getting larger for both field directions upon increasing the magnetic field amplitude. The intensity of $q_{AFM}$ was multiplied by -1 in order to transform the FT intensities into values of relative orientation of the spin-directions (green symbols). Furthermore, the background intensity at $q_{AFM}$ at 0 T was subtracted for all data points of $q_{AFM}$.

}
\end{document}